\documentclass [10pt,twoside] {article}
\usepackage{graphicx,fleqn,espcrc2}
\usepackage{epsfig}
\def\lsim{\raise0.3ex\hbox{$<$\kern-0.75em\raise-1.1ex\hbox{$\sim$}}}
\def\gsim{\raise0.3ex\hbox{$>$\kern-0.75em\raise-1.1ex\hbox{$\sim$}}}

\title{d--Au  and p--p collisions at RHIC \\ and the multichain Monte Carlo
\textsc{Dpmjet}-III 
 \thanks{Based on a contribution to a workshop "QCD at Cosmic Energies,
 Erice, Italie, Aug.29 to Sept.5 2004}
 }
\author{
 F.W. Bopp,  J. Ranft\address{Fachbereich Physik, Universit\"at
     Siegen,
 \\
        D--57068 Siegen, Germany},
         R. Engel\address{Forschungszentrum Karlsruhe, Institut f\"ur
		    Kernphysik,\\
        Postfach 3640, D--76021 Karlsruhe, Germany},
        and
					S. Roesler\address{CERN, Geneva,
					Switzerland}
					       }
\begin{document}
% +--------------------------------------------------------------------+
% |                                                                    |
% |                           TABLES.TEX                               |
% |                                                                    |
% |                     Ray F. Cowan  15-Feb-85                        |
% |                                                                    |
% |                       Princeton University                         |
% |                                                                    |
% |                     Last Revision: 17-Apr-86                       |
% |                                                                    |
% |   Macros I find handy for making tables.  See TABLEDOC TEX for     |
% |   a longer description.  The token-counting macros are straight    |
% |   from the TeXbook's "Dirty Tricks" appendix.                      |
% |                                                                    |
% +--------------------------------------------------------------------+
%
\newbox\hdbox%
\newcount\hdrows%
\newcount\multispancount%
\newcount\ncase%
\newcount\ncols% This is the number of primary text columns in the table.
\newcount\nrows%
\newcount\nspan%
\newcount\ntemp%
\newdimen\hdsize%
\newdimen\newhdsize%
\newdimen\parasize%
\newdimen\spreadwidth%
\newdimen\thicksize%
\newdimen\thinsize%
\newdimen\tablewidth%
\newif\ifcentertables%
\newif\ifendsize%
\newif\iffirstrow%
\newif\iftableinfo%
\newtoks\dbt%
\newtoks\hdtks%
\newtoks\savetks%
\newtoks\tableLETtokens%
\newtoks\tabletokens%
\newtoks\widthspec%
%
%  Book-keeping stuff--see how often these macros are called.
%
\immediate\write15{%
CP SMSG GJMSINK TEXTABLE --> TABLE MACROS V. 851121 JOB = \jobname%
}%
%
%  Turn on table diagnostics.
%
\tableinfotrue%
\catcode`\@=11%  Allows use of "@" in macro names, like PLAIN.TEX does.
\def\out#1{\immediate\write16{#1}}%  Debugging aid.  Writes #1 on the
%                                    user's terminal and in the log file.
%
%  Define the \tstrut height, depth in terms of the x_height parameter.
%
\def\tstrut{\vrule height3.1ex depth1.2ex width0pt}%
\def\and{\char`\&}%  Allows us to get an `&' in the text.  This is the
%                    same as using the PLAIN TeX macro \&.
\def\tablerule{\noalign{\hrule height\thinsize depth0pt}}%
\thicksize=1.5pt%  Default thickness for fat rules.  The user should feel
%                  free to change this to his preference.
\thinsize=0.6pt%   Default thickness for thin rules.
\def\thickrule{\noalign{\hrule height\thicksize depth0pt}}%
\def\hrulefill{\leaders\hrule\hfill}%
\def\bigrulefill{\leaders\hrule height\thicksize depth0pt \hfill}%
\def\ctr#1{\hfil\ #1\hfil}%
\def\altctr#1{\hfil #1\hfil}%
\def\vctr#1{\hfil\vbox to0pt{\vss\hbox{#1}\vss}\hfil}%
%
%  Here are things for controlling the width of the finished table.
%
\tablewidth=-\maxdimen%
\spreadwidth=-\maxdimen%
\def\tabskipglue{0pt plus 1fil minus 1fil}%
%
%  Stuff for centering or not.
%
\centertablestrue%
\def\centeredtables{%
   \centertablestrue%
}%
\def\noncenteredtables{%
   \centertablesfalse%
}%
%
%  \vctr vertically centers its argument in the row.
%
\parasize=4in%
\long\def\para#1{%  Used to make little paragraphs out of one entry.
   {%
      \vtop{%
         \hsize=\parasize%
         \baselineskip14pt%
         \lineskip1pt%
         \lineskiplimit1pt%
         \noindent #1%
         \vrule width0pt depth6pt%
      }%
   }%
}%
\gdef\ARGS{########}%  Produces the correct number of #'s in the preamble
%                      by the time eveything is expanded and \halign sees
%                      it.
\gdef\headerARGS{####}%  Same as \ARGS, but used in \header macros.
\def\@mpersand{&}%  Allows us to get alignment tab characters later
%                   when we have made the character "&" an active macro.
{\catcode`\|=13%  Make |'s locally active.
\gdef\letbarzero{\let|0}%  Globally define a macro that allows us to
%                          keep active |'s from being expanded in edef's.
\gdef\letbartab{\def|{&&}}%
\gdef\letvbbar{\let\vb|}%
%  This \def will cause active |'s read by
%                            \ruledtable to be converted into double
%                            alignment tabs.
}%  End of locally active |'s.
{\catcode`\&=4%  Make these alignment tabs.
\def\ampskip{&\omit\hfil&}%  This local macro skips a vertical rule.
\catcode`\&=13%  Now make &'s into active macros.
\let&0%  This allows us to expand \ampskip in the next \xdef without
%        attempting to expand the & and getting an "undefined control
%        sequence" error.
\xdef\letampskip{\def&{\ampskip}}%
\gdef\letnovbamp{\let\novb&\let\tab&}
%  This will cause active &'s read by
%                                   \ruledtable to be converted into
%                                   double tabs and an \omit'ted \vrule.
}%  End of locally active &'s.
\def\begintable{%  Here we make |'s and &'s active characters so we can
%                  interpret them as macros.  Note that this action is
%                  true only until we encounter the matching \endgroup
%                  token later at the end of the \ruledtable macro.
   \begingroup%
   \catcode`\|=13\letbartab\letvbbar%
   \catcode`\&=13\letampskip\letnovbamp%
   \def\multispan##1{%  We must redefine \multispan to count the number
%                       of primary columns, not physical columns.
      \omit \mscount##1%
      \multiply\mscount\tw@\advance\mscount\m@ne%
      \loop\ifnum\mscount>\@ne \sp@n\repeat%
   }%  End of \multispan macro.
   \def\|{%
      &\omit\widevline&%
   }%
   \ruledtable%  Now we call \ruledtable to do the real work.
}%  End of \begintable macro.
\long\def\ruledtable#1\endtable{%
%
%  This macro reads in the user's data entries
%  and converts them into a ruled table.
%
%  Important note:  Many macros and parameters are re-defined here, and
%  these must be kept local to the table macros to avoid conflict with
%  their use outside of tables.  This is done by the \begingroup token
%  macro \begintable and the \endgroup token at the end of
%  this macro.
%
   \offinterlineskip%  Needed to make rules touch each other.
   \tabskip 0pt%  Needed for same reason as \offinterlineskip.
   \def\widevline{\vrule width\thicksize}%  Make outer \vrule's wider.
   \def\endrow{\@mpersand\omit\hfil\crnorm\@mpersand}%
   \def\crthick{\@mpersand\crnorm\thickrule\@mpersand}%
   \def\crthickneg##1{\@mpersand\crnorm\thickrule
          \noalign{{\skip0=##1\vskip-\skip0}}\@mpersand}%
   \def\crnorule{\@mpersand\crnorm\@mpersand}%
   \def\crnoruleneg##1{\@mpersand\crnorm
          \noalign{{\skip0=##1\vskip-\skip0}}\@mpersand}%
   \let\nr=\crnorule%  A shorter abbreviation.
   \def\endtable{\@mpersand\crnorm\thickrule}%
   \let\crnorm=\cr%  Allows us to use \cr for our own purposes.
%
%  Cause user-typed \cr's to follow a row with a \tablerule.
%
   \edef\cr{\@mpersand\crnorm\tablerule\@mpersand}%
   \def\crneg##1{\@mpersand\crnorm\tablerule
          \noalign{{\skip0=##1\vskip-\skip0}}\@mpersand}%
   \let\ctneg=\crthickneg
   \let\nrneg=\crnoruleneg
   \the\tableLETtokens%  Get the user's extra \let's, if any.
%
%  Put the data entries into a token register so we can scan through them
%  and see what the user is asking us to do.
%
   \tabletokens={&#1}%  We add an extra alignment tab to the beginning
%                       of the first row to allow for the first \vrule.
%
%  Now count how many rows are in the table and return the result in
%  count register \nrows; do the same for columns, and return that
%  in register \ncols.
%
   \countROWS\tabletokens\into\nrows%
   \countCOLS\tabletokens\into\ncols%
%
%  Now do a little arithmetic to convert the number of primary columns
%  into the number of physical columns that the alignment preamble must
%  prepare for;  similarly for rows.
%
   \advance\ncols by -1%
   \divide\ncols by 2%
   \advance\nrows by 1%
%
%  Tell the user how many rows and columns we found in his data, if he
%  wants to know.
%
   \iftableinfo %
      \immediate\write16{[Nrows=\the\nrows, Ncols=\the\ncols]}%
   \fi%
%
%  Now we actually go ahead and produce the table.
%
   \ifcentertables
      \ifhmode \par\fi%  Make sure we are in vertical mode.
%     \line{% The final table comes out as an \hbox of width the \hsiz
%                    * * Replaced  at 14:32:17 on 3 Mar 1987 by JOWETT
%      since \line is usurped by LaTeX ...
      \hbox to \hsize{% The final table comes out as an \hbox of width the \hsiz
      \hss%  The final table will be centered left-to-right.
   \else %
      \hbox{%
   \fi
      \vbox{%
         \makePREAMBLE{\the\ncols}%  Generate the preamble.
         \edef\next{\preamble}%  This line and the next line force the
         \let\preamble=\next%    expansion of all \ARGS tokens into the
%                                appropriate number of #'s.
         \makeTABLE{\preamble}{\tabletokens}%  Go do the \halign here.
      }%  End of \vbox.
      \ifcentertables \hss}\else }\fi%  Finish the centering effect.
%                                       It is important that no spaces
%                                       follow the two `}' here.
%  }%  End of \line.
   \endgroup%  Return all local macros and parameters to their outside
%              values.
   \tablewidth=-\maxdimen%  Reset \tablewidth to normal.
   \spreadwidth=-\maxdimen% Same for \spreadwidth.
}%  End of macro \ruledtable.
\def\makeTABLE#1#2{%  Does an \halign for the \ruledtable macro.
   {%  Start of local parameter values.
   \let\ifmath0%     These macros would cause trouble if they were to be
   \let\header0%     expanded in the following \xdef; we \let them be
   \let\multispan0%  equal to a digit, because digits can't be expanded.
%
%  Set up the width specification here.
%
   \ncase=0%
   \ifdim\tablewidth>-\maxdimen \ncase=1\fi%
   \ifdim\spreadwidth>-\maxdimen \ncase=2\fi%
   \relax%  This \relax is absolutely necessary, without it the following
%           \ifcase will always take \ncase=0.
%
   \ifcase\ncase %
      \widthspec={}%
   \or %
      \widthspec=\expandafter{\expandafter t\expandafter o%
                 \the\tablewidth}%
   \else %
      \widthspec=\expandafter{\expandafter s\expandafter p\expandafter r%
                 \expandafter e\expandafter a\expandafter d%
                 \the\spreadwidth}%
   \fi %
%\out{Widthspec=[\the\widthspec]}%
%\out{Preamble=[\preamble]}%
   \xdef\next{%  We must force the preamble to be expanded BEFORE the
      \halign\the\widthspec{%
%        \halign is done;  this \edef\next{...}\next construction
%                does the trick.
      #1%  This is the preamble text.
      \noalign{\hrule height\thicksize depth0pt}%  Makes the top \hrule.
      \the#2\endtable%  This is the main body.
%
%     \noalign{\hrule height0.7pt depth0pt}%  Makes the last \hrule.
      }%  End of \halign.
   }%  End of \next.
   }%  End of local values.
   \next%  This \next must be outside of the local values, because now
%          we want those troublesome macros in the \let's above to have
%          their normal actions.
}%  End of macro \makeTABLE.
\def\makePREAMBLE#1{%  This macro generates the necessary preamble for a
%                      ruled table with #1 primary columns.
%                      (Primary columns means the number of columns NOT
%                       counting those used for vertical rules.)
   \ncols=#1%  Get the number of columns desired.
   \begingroup%  Start local parameter definitions.
   \let\ARGS=0%  This is the key to the whole thing; it prevents \ARGS
%                from being expanded in the following \edef's.
   \edef\xtp{\widevline\ARGS\tabskip\tabskipglue%
   &\ctr{\ARGS}\tstrut}%  A 1-column preamble.  Gets the sizing right.
   \advance\ncols by -1%  One column has been generated; decrement the
%                         counter.
   \loop%  Append as many further columns as needed to the preamble.
      \ifnum\ncols>0 %
      \advance\ncols by -1%
      \edef\xtp{\xtp&\vrule width\thinsize\ARGS&\ctr{\ARGS}}%
   \repeat
   \xdef\preamble{\xtp&\widevline\ARGS\tabskip0pt%
   \crnorm}%  Adds the last \vrule.
   \endgroup%  End of local parameters.
}%  End of macro \makePREAMBLE.
\def\countROWS#1\into#2{%  This counts the number of rows in #1 by
%                          looking for control sequences that end a row,
%                          e.g., \cr, \crthick, etc., and puts the result
%                          into count register #2.
   \let\countREGISTER=#2%
   \countREGISTER=0%
%  \out{In countROWS:  tokens are [\the#1]}%
   \expandafter\ROWcount\the#1\endcount%
}%
\def\ROWcount{%
   \afterassignment\subROWcount\let\next= %
}%
\def\subROWcount{%
%  \out{In subROWcount:  next is [\meaning\next]}%  Debugging aid.
   \ifx\next\endcount %
      \let\next=\relax%
   \else%
      \ncase=0%
      \ifx\next\cr %
         \global\advance\countREGISTER by 1%
         \ncase=0%
      \fi%
      \ifx\next\endrow %
         \global\advance\countREGISTER by 1%
         \ncase=0%
      \fi%
      \ifx\next\crthick %
         \global\advance\countREGISTER by 1%
         \ncase=0%
      \fi%
      \ifx\next\crnorule %
         \global\advance\countREGISTER by 1%
         \ncase=0%
      \fi%
      \ifx\next\crthickneg %
         \global\advance\countREGISTER by 1%
         \ncase=0%
      \fi%
      \ifx\next\crnoruleneg %
         \global\advance\countREGISTER by 1%
         \ncase=0%
      \fi%
      \ifx\next\crneg %
         \global\advance\countREGISTER by 1%
         \ncase=0%
      \fi%
      \ifx\next\header %
%     \out{In subROWcount:  next=header, ncase set=1}%
         \ncase=1%
      \fi%
%     \out{In subROWcount:  ncase is [\the\ncase]}%
      \relax%
      \ifcase\ncase %
         \let\next\ROWcount%
%        \out{subROWcount---> ncase=\the\ncase}%
      \or %
         \let\next\argROWskip%
%        \out{subROWcount---> ncase=\the\ncase}%
      \else %
      \fi%
   \fi%
%  \out{subROWcount---> NEXT=\meaning\next}%
   \next%
}%  End of macro \subROWcount.
\def\counthdROWS#1\into#2{%
\dvr{10}%
   \let\countREGISTER=#2%
   \countREGISTER=0%
\dvr{11}%
%  \out{In counthdROWS:  tokens are [\the#1]}%
\dvr{13}%
   \expandafter\hdROWcount\the#1\endcount%
\dvr{12}%
}%
\def\hdROWcount{%
   \afterassignment\subhdROWcount\let\next= %
}%
\def\subhdROWcount{%
%\out{In subhdROWcount:  next is [\meaning\next]}%
   \ifx\next\endcount %
      \let\next=\relax%
   \else%
      \ncase=0%
      \ifx\next\cr %
         \global\advance\countREGISTER by 1%
         \ncase=0%
      \fi%
      \ifx\next\endrow %
         \global\advance\countREGISTER by 1%
         \ncase=0%
      \fi%
      \ifx\next\crthick %
         \global\advance\countREGISTER by 1%
         \ncase=0%
      \fi%
      \ifx\next\crnorule %
         \global\advance\countREGISTER by 1%
         \ncase=0%
      \fi%
      \ifx\next\header %
%\out{In subhdROWcount:  next=header, ncase set=1}%
         \ncase=1%
      \fi%
%\out{In subhdROWcount:  ncase is [\the\ncase]}%
\relax%
      \ifcase\ncase %
         \let\next\hdROWcount%
%\out{subhdROWcount---> ncase=\the\ncase}%
      \or%
         \let\next\arghdROWskip%
%\out{subhdROWcount---> ncase=\the\ncase}%
      \else %
      \fi%
   \fi%
%\out{subhdROWcount---> NEXT=\meaning\next}%
   \next%
}%
{\catcode`\|=13\letbartab
\gdef\countCOLS#1\into#2{%
%  \out{In countCOLS:  tokens are [\the#1]}
   \let\countREGISTER=#2%
   \global\countREGISTER=0%
   \global\multispancount=0%
   \global\firstrowtrue
   \expandafter\COLcount\the#1\endcount%
   \global\advance\countREGISTER by 3%
   \global\advance\countREGISTER by -\multispancount
%  \out{countCOLS-->[\the\countREGISTER]}
}%
\gdef\COLcount{%
   \afterassignment\subCOLcount\let\next= %
}%
{\catcode`\&=13%
\gdef\subCOLcount{%
%\out{In subCOLcount: next is [\meaning\next]}
   \ifx\next\endcount %
      \let\next=\relax%
   \else%
      \ncase=0%
      \iffirstrow
         \ifx\next& %
            \global\advance\countREGISTER by 2%
            \ncase=0%
         \fi%
         \ifx\next\span %
            \global\advance\countREGISTER by 1%
            \ncase=0%
         \fi%
         \ifx\next| %
            \global\advance\countREGISTER by 2%
            \ncase=0%
         \fi
         \ifx\next\|
            \global\advance\countREGISTER by 2%
            \ncase=0%
         \fi
         \ifx\next\multispan
            \ncase=1%
            \global\advance\multispancount by 1%
         \fi
         \ifx\next\header
            \ncase=2%
         \fi
         \ifx\next\cr       \global\firstrowfalse \fi
         \ifx\next\endrow   \global\firstrowfalse \fi
         \ifx\next\crthick  \global\firstrowfalse \fi
         \ifx\next\crnorule \global\firstrowfalse \fi
         \ifx\next\crnoruleneg \global\firstrowfalse \fi
         \ifx\next\crthickneg  \global\firstrowfalse \fi
         \ifx\next\crneg       \global\firstrowfalse \fi
      \fi%  End of \iffirstrow.
\relax%\out{subCOL-->  ncase=[\the\ncase]}
% \out{subCOL-->  next=\meaning\next}
      \ifcase\ncase %
         \let\next\COLcount%
      \or %
         \let\next\spancount%
      \or %
         \let\next\argCOLskip%
      \else %
      \fi %
   \fi%
%  \out{subCOL-->  countREGISTER=[\the\countREGISTER]}
   \next%
}%
\gdef\argROWskip#1{%
%  Deletes the next balanced, undelimited argument from a
%                 token list.
% \out{---> Entering argROWskip <---}
% \out{In argROWskip:  deleted arg is [#1]}%
   \let\next\ROWcount \next%
}%  End of macro \argskip.
\gdef\arghdROWskip#1{%
%  Deletes the next balanced, undelimited argument from a
%                 token list.
% \out{---> Entering arghdROWskip <---}
% \out{In arghdROWskip:  deleted arg is [#1]}%
   \let\next\ROWcount \next%
}%  End of macro \arghdROWskip.
\gdef\argCOLskip#1{%
%  Deletes the next balanced, undelimited argument from a
%                 token list.
% \out{---> Entering argCOLskip <---}
% \out{In argCOLskip:  deleted arg is [#1]}%
   \let\next\COLcount \next%
}%  End of macro \argskip.
}%  End of active &'s.
}%  End of active |'s.
\def\spancount#1{%\out{spancount--->\meaning#1}
   \nspan=#1\multiply\nspan by 2\advance\nspan by -1%
   \global\advance \countREGISTER by \nspan
%  \out{number spancount--->\the\nspan; \the\countREGISTER}
   \let\next\COLcount \next}%
\def\dvr#1{\relax}%
% \omit\hfil%
% \parindent=0pt\hsize=1.1in\valign{%
% \vfil#\vfil&\vfil#\vfil\cr\hfil\hbox{\ Added to\ }\hfil&%
% \hfil\hbox{\ empty events\ }\hfil\cr}\hfil%
\def\header#1{%
\dvr{1}{\let\cr=\@mpersand%
\hdtks={#1}%
%\out{In header:  hdtks=[\the\hdtks]}%
\counthdROWS\hdtks\into\hdrows%
\advance\hdrows by 1%
\ifnum\hdrows=0 \hdrows=1 \fi%
%\out{In header:  Nhdrows=[\the\hdrows]}%
\dvr{5}\makehdPREAMBLE{\the\hdrows}%
%\out{In header:  headerpreamble=[\headerpreamble]}%
\dvr{6}\getHDdimen{#1}%
%\out{In header:  hdsize=[\the\hdsize]}%
%\striplastCR{#1}%
{\parindent=0pt\hsize=\hdsize{\let\ifmath0%
\xdef\next{\valign{\headerpreamble #1\crnorm}}}\dvr{7}\next\dvr{8}%
}%
}\dvr{2}}%  End of macro \header.
\def\makehdPREAMBLE#1{%This macro generates the necessary preamble for a
\dvr{3}%
%                      ruled table with \ncols primary columns.
%                      (Primary columns means the number of columns NOT
%                       counting those used for vertical rules.
\hdrows=#1%  Get the number of columns desired.
{%  Start local parameter definitions.
\let\headerARGS=0%
%  This is the key to the whole thing; it prevents \ARGS
\let\cr=\crnorm%
%                from being expanded in the followin \edef's.
\edef\xtp{\vfil\hfil\hbox{\headerARGS}\hfil\vfil}%
\advance\hdrows by -1%  One row has been generated; decrement the
%                         counter.
\loop%  Append as many further rows as needed to the preamble.
\ifnum\hdrows>0%
\advance\hdrows by -1%
\edef\xtp{\xtp&\vfil\hfil\hbox{\headerARGS}\hfil\vfil}%
\repeat%
\xdef\headerpreamble{\xtp\crcr}%
}%  End of local parameters.
\dvr{4}}%  End of \makehdPREAMBLE.
\def\getHDdimen#1{%
%\out{In getHDdimen:  Arg 1=[#1]}%
\hdsize=0pt%
\getsize#1\cr\end\cr%
}%  End of macro getHDdimen.
\def\getsize#1\cr{%
%\out{In getsize:  Arg 1=[#1]}%
%  Here we have to check arg#1 and see if the first token in #1 is an
%    \end; if so, we stop, else we check the width of arg#1.
%  We recall that each arg#1 will be terminated with a \cr token.
\endsizefalse\savetks={#1}%
%\out{In getsize:  the savetks = [\the\savetks]}%
\expandafter\lookend\the\savetks\cr%
%\out{In getsize:  ifendsize = [\meaning\ifendsize]}%
\relax \ifendsize \let\next\relax \else%
\setbox\hdbox=\hbox{#1}\newhdsize=1.0\wd\hdbox%
\ifdim\newhdsize>\hdsize \hdsize=\newhdsize \fi%
%\out{In getsize:  hdsize=[\the\hdsize]}%
%\out{In getsize:  newhdsize=[\the\newhdsize]}%
\let\next\getsize \fi%
\next%
}%
\def\lookend{\afterassignment\sublookend\let\looknext= }%
\def\sublookend{\relax%
%\out{In sublookend:  looknext = [\looknext]}%
\ifx\looknext\cr %
%\out{In sublooknext:  looknext=cr}%
\let\looknext\relax \else %
%\out{In sublooknext:  looknext/=cr}%
   \relax
   \ifx\looknext\end \global\endsizetrue \fi%
   \let\looknext=\lookend%
    \fi \looknext%
}%
%
%  Allow the user to make his own names for crthick, etc.
%
\def\tablelet#1{%
   \tableLETtokens=\expandafter{\the\tableLETtokens #1}%
}%
\catcode`\@=12%  Change @'s back to their normal category code.

\begin{abstract}
%{\bf Abstract}

In this paper we compare systematically 
 the two-component
Dual Parton Model (DPM) event generator \textsc{Dpmjet}-III to d-Au and p--p 
data from
RHIC. In this process we are able to improve the model. The need for
fusion of chains and a recalibration of the model to obtain collision
scaling in h-A and d-A collisions was found already in previous
comparisons. Here, comparing to transverse momentum distributions of
identified charged hadrons we find also the need to modify the
transverse momentum distributions in the decay of hadronic strings,
the basic building blocks of the model on soft hadronic collisions. 

\end{abstract}
\maketitle

\vspace*{-10mm}
\section{Introduction}

Hadronic collisions at high energies involve the production of particles
with low transverse momenta, the so-called \textit{soft} multiparticle
production. 
The theoretical tools available at present are not sufficient
to understand this feature from QCD alone and phenomenological models
are typically applied in addition to perturbative QCD. The Dual Parton
Model (DPM) \cite{Capella94a} is such a {phenomenological}
model and its fundamental ideas are presently the basis of many of
the Monte Carlo implementations of soft interactions.

The soft component of the Dual Parton Model contains
some features to be determined from multi--particle production data.
This is  so for the fragmentation of chains. But it also
concerns the general event structure.
Already in Ref. \cite{posterpap04} we have used data from RHIC to tune
some of the properties of the soft particle production model contained
in \textsc{Dpmjet}--III. 
This process will be continued here using mainly
RHIC date on d-Au and p--p collisions.

 The \textsc{Dpmjet}-III code system \cite{Roesler20001,Roesler20002}
 is a Monte Carlo event 
generator implementing Gribov--Glauber theory for collisions involving
nuclei, for all elementary collisions it uses the  DPM and (LO) 
perturbative QCD as implemented in   
 \textsc{Phojet}1.12 \cite{Engel95a,Engel95d}.

 \textsc{Dpmjet}-III is unique in its wide range of application
 simulating hadron-hadron,
hadron-nucleus,
nucleus-nucleus, photon-hadron, photon-photon and 
photon-nucleus interactions
from a few GeV up to  cosmic ray energies.

The properties of  \textsc{Dpmjet}--III were recently summarized in
Ref. \cite{posterpap04}. For a more datailled description of
\textsc{Dpmjet}--III we refer to this paper and the literature quoted
there.

\section{Comparing the original \textsc{Dpmjet}--III with RHIC data}

In Ref. \cite{posterpap04} we found an        
 excellent agreement of \textsc{Dpmjet}--III 
 up to transverse momenta of about
$p_{\perp}$ = 10 GeV/c  to 
 $\pi^0$ transverse momentum distribution
in p--p collisions at $\sqrt s$ = 200 GeV from PHENIX \cite{PHENIX02a}.

In the same paper, comparing \textsc{Dpmjet}--III with multiplicity
data from RHIC in central Au--Au collisions we found a problem which let us
 introduce percolation and chain fusion into \textsc{Dpmjet}--III.
The groups at Lisboa \cite{Dias2000a,Dias2000b,Dias2001a} 
and Santiago de Compostela \cite{Braun2002a} were the
first to point out, that the multiplicities measured at RHIC are
significantly lower than predicted by conventional multi--string
models. A new process is needed to lower the multiplicity in
situations with a very high density of produced hadrons like in central
nucleus--nucleus collisions. 
{The
percolation process, which leads with increasing density to more and
more fusion of strings, is one such mechanism} \cite{Braun99,Braun2000a}.
\vspace{-4mm}
\section{Pseudorapidity distributions in  d-Au and p--p collisions and 
{\sc Dpmjet}-III
 }
       
\begin{figure}[thb]
\vspace{-11mm}
\begin{center}
\includegraphics[height=7cm,width=8cm]{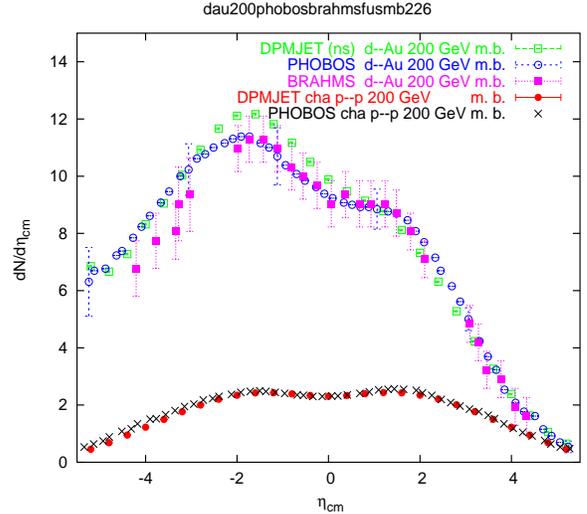}
\end{center}
\vspace{-13mm}

\caption{Pseudorapidity distribution of charged hadrons produced in minimum
bias $\sqrt{s} = 200$ GeV d--Au and p--p collisions. The results of \textsc{Dpmjet}
with  fusion of chains are compared
to preliminary experimental data from the BRAHMS--Collaboration\cite{BRAHMSdau} and
the PHOBOS--Collaboration\cite{PHOBOSdau,PHOBOSdaupp}. At some pseudorapidity values the systematic 
PHOBOS--errors as estimated
by the experimental collaboration are given.
}

\vspace{-9mm}
\label{fig:etacmdau200}
\end{figure}

Pseudorapidity distribution of charged hadrons produced in minimum bias $\sqrt{s}$
= 200 GeV d--Au collisions were measured at RHIC by the 
PHOBOS--Collaboration\cite{PHOBOSdau} and by the BRAHMS Collaboration\cite{BRAHMSdau}. 
In Fig.\ref{fig:etacmdau200} we compare the preliminary PHOBOS
data and the preliminary BRAHMS data to \textsc{Dpmjet}-III calculations. 
Using \textsc{Dpmjet}-III
without  fusion of chains we find
the \textsc{Dpmjet} distribution above the experimental data outside
the systematic errors. Using \textsc{Dpmjet}-III {with
fusion} of chains we find the \textsc{Dpmjet} distribution within
the systematic errors of both experiments. 

In \cite{PHOBOSdaupp} the PHOBOS Collaboration also presents preliminary data on
the pseudorapidity distribution of charged hadrons in p--p collisions  at
$\sqrt{s}$ = 200 GeV. In Fig.\ref{fig:etacmdau200} we compare
\textsc{Dpmjet}-III  also with these data and find an excellent
agreement.
 
\begin{figure}[thb]
\vspace{-11mm}
\begin{center}
\includegraphics[height=7cm,width=8cm]{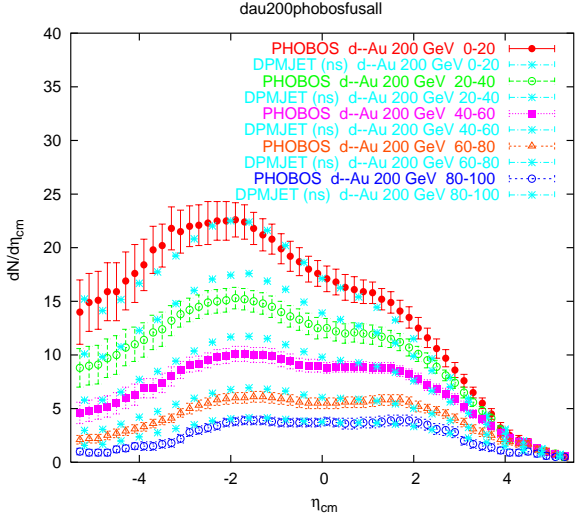}
\end{center}
\vspace{-13mm}

\caption{Pseudorapidity distribution of charged hadrons produced in
non-single-diffractive (ns) 
 $\sqrt{s}$ = 200 GeV d--Au collisions with different centralities. 
 The results of \textsc{Dpmjet}
with  fusion of chains are compared
to preliminary experimental data from the PHOBOS--Collaboration\cite{PHOBOSdauall}.
 }

\vspace{-7mm}
\label{fig:etacmdau200all}
\end{figure}

{Pseudorapidity distributions of char\-ged ha\-drons in d--Au collisions with
specified centralities were given by the BRAHMS--Collaboration\cite{BRAHMSdau} 
and by the PHOBOS--Collaboration\cite{PHOBOSdauall}.
In Fig.\ref{fig:etacmdau200all} we compare the preliminary PHOBOS--data with results
from \textsc{Dpmjet}. In the most central collisions we find an
excellent agreement. In  \textsc{Dpmjet}-III we know the impact
parameter of the collision and we determine the centrality according to
the impact parameter distribution. This might give results slightly
different from the PHOBOS centrality determination, also the the number
of participants determined by PHOBOS and obtained from
\textsc{Dpmjet}-III differs for centralities between 20 and 80 \% , see
Table 1.}

\vspace{15mm}
\noindent { Table 1. Numbers of participants in d--Au collisions with
different centralities. PHOBOS numbers taken from
\protect\cite{PHOBOSdauall}}

\vspace{3mm}
{\small \begintable
{\tiny Centrality \%}|{ source}|~$N_{part}$~|~$N_{part}$(Au)~| ~$N_{part}$(d)~\cr
 0--20 \%|PHOBOS |15.5 |13.5 |2.0  \cr
  |\textsc{Dpmjet}|15.4 |13.5 |2.0  \cr
 20--40 \%| PHOBOS|10.8  |8.9 | 1.9 \cr
  |\textsc{Dpmjet} |11.7 |9.8 |1.94  \cr
 40--60 \%|PHOBOS | 7.2 |5.4 |1.7  \cr
  |\textsc{Dpmjet} | 7.8 |6.0 |1.76  \cr
 60--80 \%|PHOBOS |4.2  |2.9 |1.4  \cr
  |\textsc{Dpmjet} | 4.7 |3.24 |1.41  \cr
 80--100 \%|PHOBOS | 2.7 |1.6 |1.1  \cr
  |\textsc{Dpmjet} |2.9 |1.8 |1.1  \endtable }

 %\vspace{5mm}

\section{Transverse momentum and transverse energy 
distributions of charged hadrons 
in  d-Au collisions and 
{\sc Dpmjet}-III
 }
       
Let us repeat an important {\sc Dpmjet} modification introduced
already in our previous paper \cite{posterpap04}.
Several RHIC experiments (see for instance 
 \cite{PHENIX02dau}) found in d-Au collisions at large $p_{\perp}$ a 
 nearly perfect collision scaling for $\pi^0$ production. 
Collision scaling means $R_{AA} \approx 1.0$ where 
the $R_{AA}$ ratios are defined as follows:
\begin{equation}
R_{AA} =
\frac{\frac{d^2}{dp_{\perp}d\eta}N^{A-A}}{N^{A-A}_{binary} \cdot
\frac{d^2}{dp_{\perp}d\eta}N^{N-N}}
\end{equation}
Here $N^{A-A}_{binary}$ is the number of binary Glauber collisions in
the nucleus--nucleus collision A--A. 

 \textsc{Dpmjet}--III  in its original form gave for $\pi^0$ production in
 d+Au collisions strong deviations from collision scaling ($R_{AA}
 \approx$ 0.5 at large $p_{\perp}$). 
 The reason is understood\cite{posterpap04}.
 Using a modified
iteration procedure it was possible in  \cite{posterpap04} 
to obtain a nearly perfect collision scaling. 

\begin{figure}[thb]
\vspace{-4mm}
\begin{center}
\includegraphics[height=7cm,width=8cm]{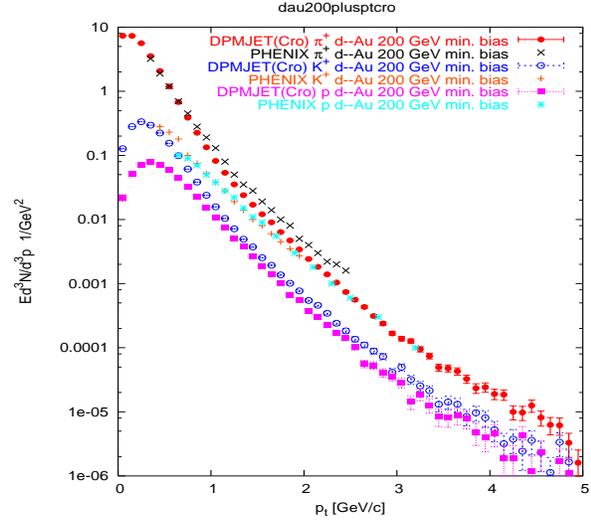}
\end{center}
\vspace{-13mm}

\caption{Transverse momentum distributions of positively charged hadrons
in minimum bias d--Au collisions. Compared are the preliminary data from the
PHENIX--Collaboration\cite{Matathias04} to the results of the original
\textsc{Dpmjet}--III.
 }

\vspace{-7mm}
\label{fig:dauplusptorig}
\end{figure}

 Having shown the collision scaling of the model in \cite{posterpap04}
 we will here compare the model directly with the measured transverse
 momentum and transverse energy distributions.
Transverse momentum distributions in minimum bias d--Au collisions were
measured by the PHENIX--Collaboration\cite{Matathias04} and by the
STAR--Collaboration\cite{Adams03}. In Fig.\ref{fig:dauplusptorig} we
compare the preliminary PHENIX--data for positively charged hadrons
with \textsc{Dpmjet}--III as described so far.

For charged Pion production we find in Fig.\ref{fig:dauplusptorig} quite
a good agreement between \textsc{Dpmjet}--III and the preliminary PHENIX--data. This
agreement is similar to the good agreement found already in
Ref. \cite{posterpap04} for $\pi^0$ production in p--p and d--Au
collisions. However, it seems, that the \textsc{Dpmjet}--III model fails
to describe the $p_{\perp}$ distributions of heavier hadrons like Kaons
and protons. The prediction is slightly too steep.
This is the first time that we are able to compare the
\textsc{Dpmjet}--III model to $p_{\perp}$ distributions  of identified
hadrons like K$^+$, K$^-$, protons  and antiprotons. In the past usually
only data for all charged hadrons were available. Therefore, it is not
surprising to find problems.

The reason for this disagreement is the following:

\textsc{Dpmjet} uses the
{\sc Pythia} code  \cite{Sjostrand01a} for the fragmentation of all chains
(hard, QCD based chains as well as soft chains). {\sc Pythia} selects in the
PYPTDI function the transverse momenta of (for instance) $q-\bar q$
pairs (to become hadrons finally) from independent Gaussians in the
$p_{\perp x}$ and $p_{\perp y}$ components of $p_{\perp}$. {\sc Pythia} 
was extensively tested in processes with a significant hard component,
for the fragmentation of hard
chains this choice is reasonable, but the fragmentation of the soft
component is not well determined.

 \begin{figure}[thb]
\vspace{-11mm}
\begin{center}
\includegraphics[height=7cm,width=8cm]{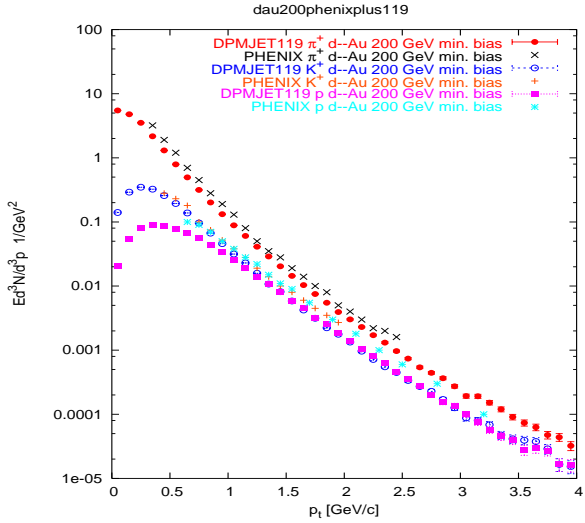}
\end{center}
\vspace{-13mm}

\caption{Transverse momentum distributions of positively charged hadrons
in minimum bias d--Au collisions. Compared are the preliminary data from the
PHENIX--Collaboration\cite{Matathias04} to the results of
\textsc{Dpmjet}--III  with modified transverse momentum distribution in
hadronic soft chain decay as described in the paper.
 }

\vspace{-7mm}
\label{fig:daupluspt}
\end{figure} 

Otherwise, it is known since very many years (given already in 1968 in
Ref. \cite{Hagedornranft}, but this might even not be the first
Reference) that the transverse momentum distribution 
for hadrons with mass m in soft hadronic
collisions is well described by an exponential in the transverse energy
\begin{equation}
\exp(-m_{\perp} /T),~~~~ m_{\perp} = \sqrt{p_{\perp}^2 + m^2}.
\end{equation}
 We here chose a similar parametrization and replace the
 Gaussian in PYPTDI of {\sc Pythia} by the
 distribution
\begin{equation}
 \exp(-\sqrt{p_{\perp}^2+m_x^2}/\sigma),~~~~m_x = 0.33 GeV/c.
\end{equation} After such a change all {\sc Pythia} parameter 
which relate to the fragmentation
have to be reoptimized.

\begin{figure}[thb]
\vspace{-6mm}
\begin{center}
\includegraphics[height=7cm,width=8cm]{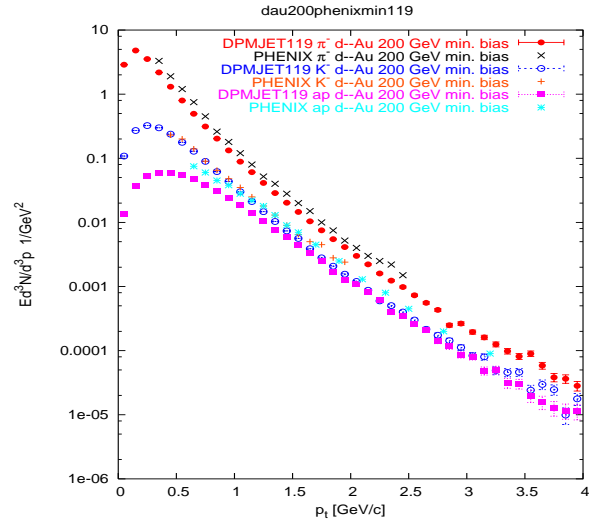}
\end{center}
\vspace{-13mm}

\caption{Transverse momentum distributions of negatively charged hadrons
in minimum bias d--Au collisions. Compared are the preliminary data from the
PHENIX--Collaboration\cite{Matathias04} to the results of
\textsc{Dpmjet}--III  with modified transverse momentum distribution in
hadronic soft chain decay as described in the paper.
 }

\vspace{-7mm}
\label{fig:dauminpt}
\end{figure}

With this modification we compare \textsc{Dpmjet}--III again to the
PHENIX--transverse momentum distributions \cite{Matathias04} in
Fig.\ref{fig:daupluspt} for positively charged hadrons and in
Fig.\ref{fig:dauminpt} for negatively charged hadrons. To be definitive:
we use \textsc{Dpmjet}--III with (i) chain fusion  \cite{posterpap04},
(ii) anomalous baryon stopping
as described in Ref. \cite{Ranft20003},
(iii) the modified iteration procedure to obtain
collision scaling \cite{posterpap04} and (iv) the changed 
transverse momentum
distribution in soft chain decay as described above. We find in both
Figures a much improved  agreement to the preliminary PHENIX--data
\cite{Matathias04}.
%For $K^+$ and $K^-$ the
%model needs further improvements. 
At sufficiently large $p_{\perp}$ the $p_{\perp}$
distributions of Pions, Kaons and baryons have the same exponential
slope.

\begin{figure}[thb]
\vspace{-6mm}
\begin{center}
\includegraphics[height=7cm,width=8cm]{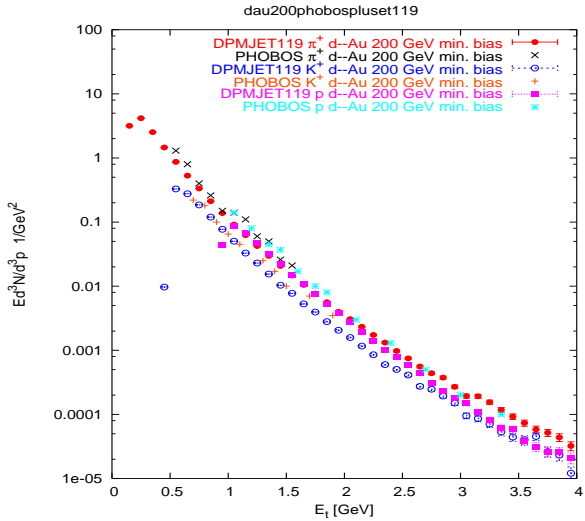}
\end{center}
\vspace{-13mm}

\caption{Transverse energy distributions of positively charged hadrons
in minimum bias d--Au collisions. Compared are the preliminary data from the
PHOBOS--Collaboration\cite{Veres04} to the results of
\textsc{Dpmjet}--III  with modified transverse momentum distribution in
hadronic soft chain decay as described in the paper.
 }

\vspace{-7mm}
\label{fig:daupluset}
\end{figure}

\begin{figure}[thb]
\vspace{-11mm}
\begin{center}
\includegraphics[height=7cm,width=8cm]{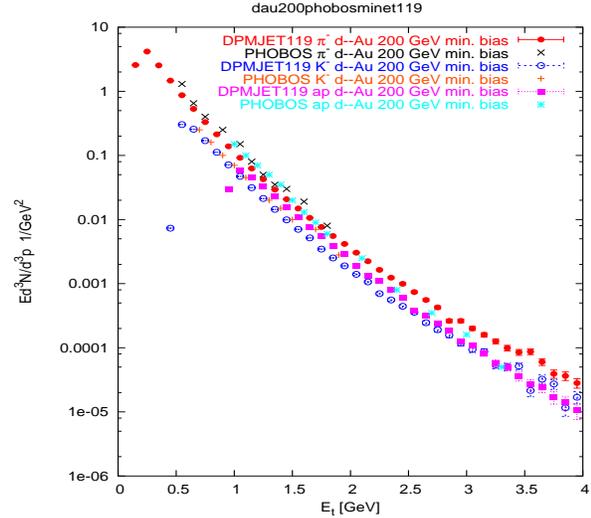}
\end{center}
\vspace{-13mm}

\caption{Transverse energy distributions of negatively charged hadrons
in minimum bias d--Au collisions. Compared are the preliminary data from the
PHOBOS--Collaboration\cite{Veres04} to the results of
\textsc{Dpmjet}--III  with modified transverse momentum distribution in
hadronic soft chain decay as described in the paper.
 }

\vspace{-7mm}
\label{fig:dauminet}
\end{figure}

As a further check we present in Figs. \ref{fig:daupluset} and
\ref{fig:dauminet} the comparison of the modified \textsc{Dpmjet}--III
with tansverse energy distributions of identified charged hadrons as
measured by the PHOBOS--Collaboration\cite{Veres04}. Before the
modification there were discrepancies between the Kaon and baryon
distributions similar to the ones in Fig.\ref{fig:dauplusptorig}. The
agreement  
in Figs. \ref{fig:daupluset} and \ref{fig:dauminet} is now much
better. 
%For $K^+$ and $K^-$ the normalization of the
%\textsc{Dpmjet}--III distributions is too low.

\begin{figure}[thb]
\vspace{-11mm}
\begin{center}
\includegraphics[height=7cm,width=8cm]{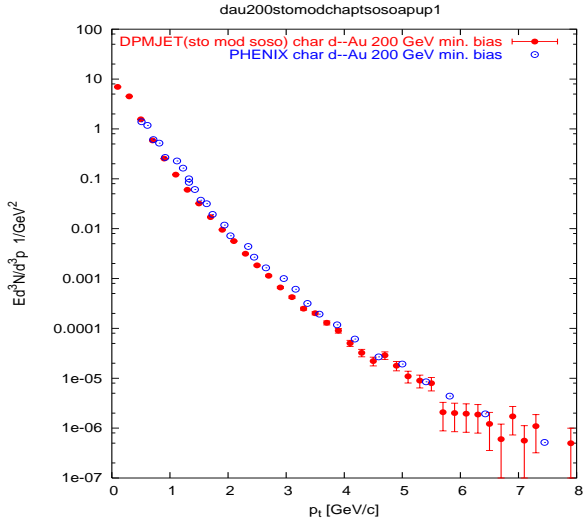}
\end{center}
\vspace{-13mm}

\caption{Transverse momentum distributions of  all charged hadrons
in minimum bias d--Au collisions. Compared are the preliminary data from the
PHENIX--Collaboration\cite{PHENIX02dau} to the results of
\textsc{Dpmjet}--III  with modified transverse momentum distribution in
hadronic soft chain decay as described in the paper.
 }

\vspace{-7mm}
\label{fig:dauchapt}
\end{figure}

Finally in Fig.\ref{fig:dauchapt} we compare the modified
\textsc{Dpmjet}--III with the transverse momentum distribution of all
charged hadrons as measured by the PHENIX--Collaboration\cite{PHENIX02dau}. 
We find a reasonable agreement.

 \begin{figure}[thb]
\vspace*{-22mm}
\begin{center}
\includegraphics[height=6cm,width=8cm]{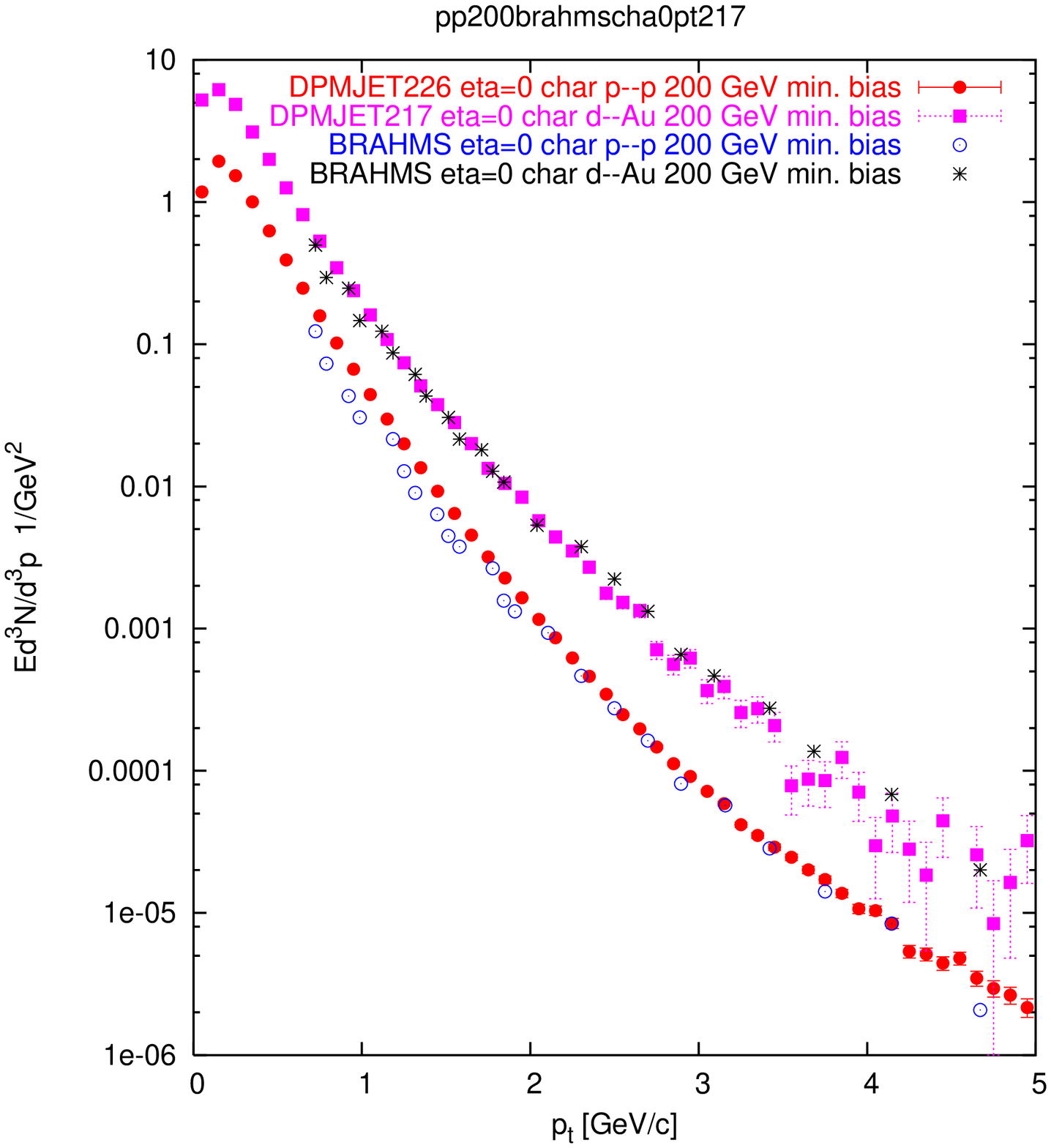}
\includegraphics[height=6cm,width=8cm]{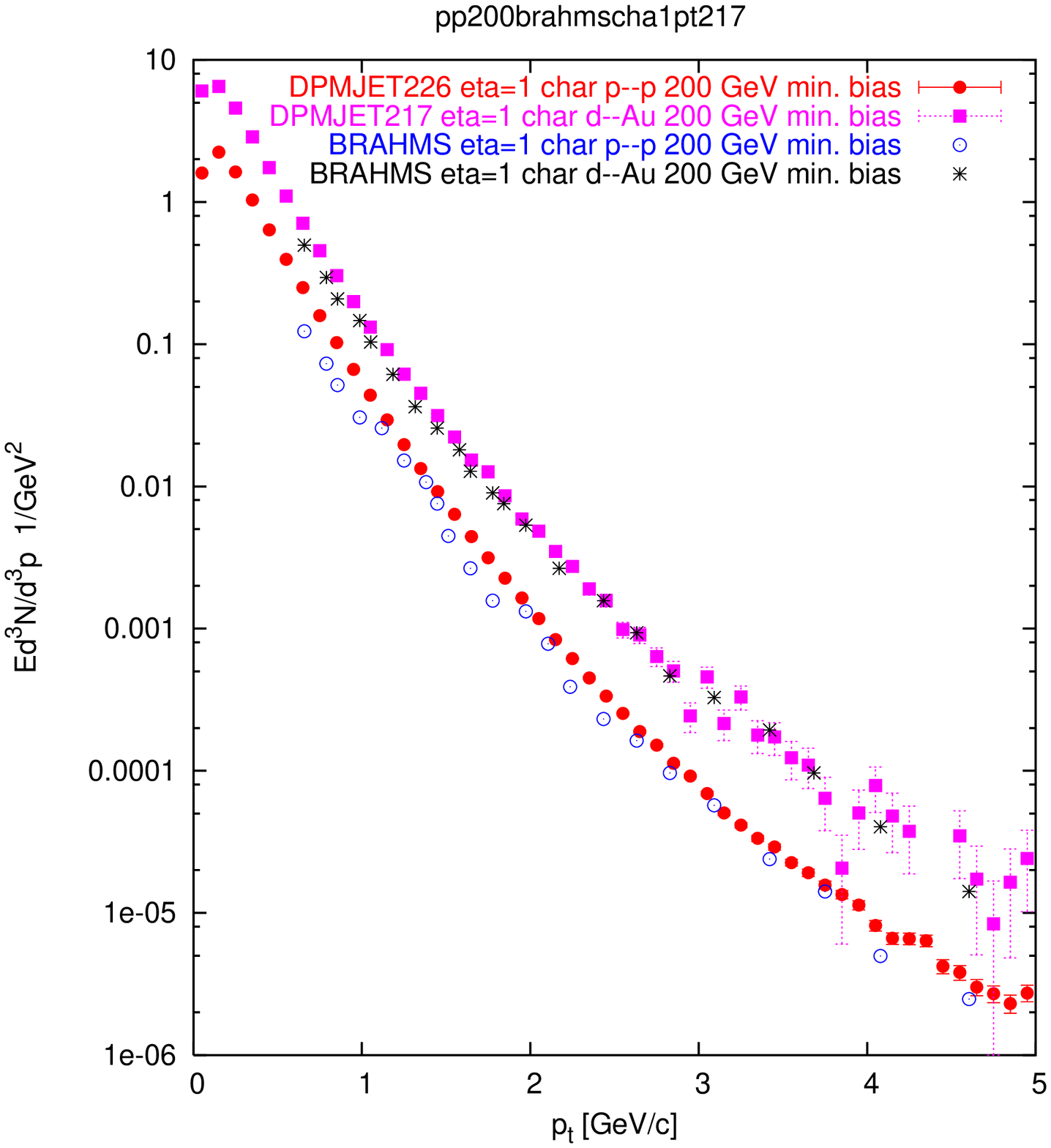}
\includegraphics[height=6cm,width=8cm]{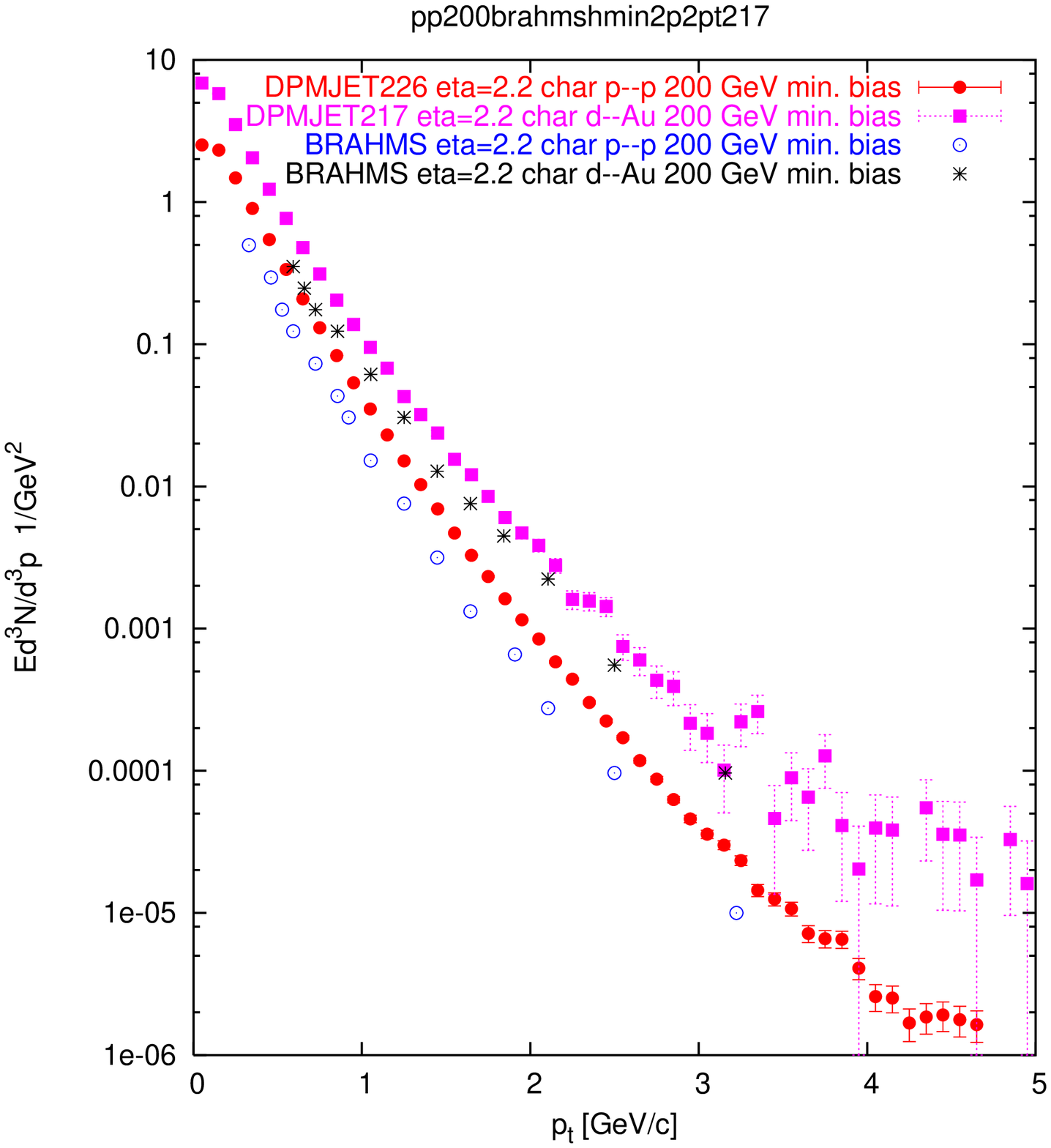}
\end{center}
\vspace{-13mm}

\
\caption{Transverse momentum distributions of  all charged hadrons
at $\eta_{cm}$ = 0, at $\eta_{cm}$ = 1, and at $\eta_{cm}$ = 2.2
in minimum bias p--p and d--Au collisions. See caption Fig.\ref{fig:daucha3p2pt}
 }

\vspace*{-7mm}
\label{fig:daucha0pt}
\label{fig:daucha1pt}
\label{fig:daucha2p2pt}
\end{figure}

\section{Transverse momentum  
distributions of charged hadrons at different pseudorapidities   
 }

\begin{figure}[thb]
\vspace{-11mm}
\begin{center}
\includegraphics[height=7cm,width=8cm]{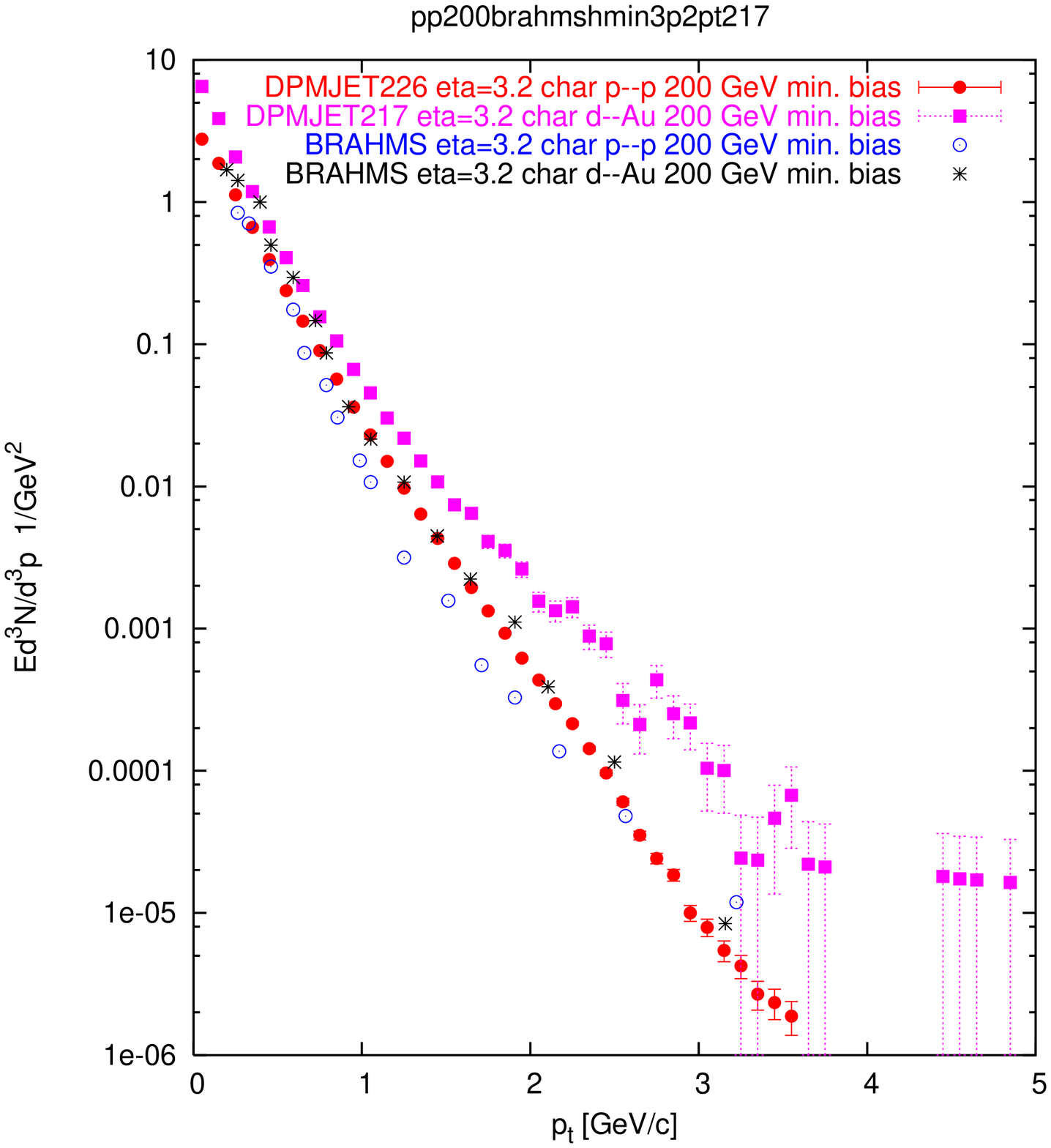}
\end{center}
\vspace{-13mm}

\caption{Transverse momentum distributions of  
all negatively charged hadrons
at $\eta_{cm}$ = 3.2
in minimum bias p--p and d--Au collisions. 
Compared are the preliminary data from the
BRAHMS--Collaboration\cite{BRAHMSdaupp} to the results of
\textsc{Dpmjet}--III  with modified transverse momentum distribution in
hadronic soft chain decay as described in the paper.
 }

\vspace{-7mm}
\label{fig:daucha3p2pt}
\end{figure}

Transverse momentum distributions of all charged hadrons and of
all negatively charged hadrons at different pseudorapidities were
measured by the  
BRAHMS--Collaboration\cite{BRAHMSdaupp} in d--Au and p--p collisions.
The interesting point in these measurements is the gradual change of the
shape of the $p_{\perp}$ distribution with changing pseudorapidity and
the corresponding evolution of the nuclear modification factors.
In Figs. \ref{fig:daucha0pt}
and \ref{fig:daucha3p2pt} we compare the modified
\textsc{Dpmjet}--III as described in this paper with the preliminary d--Au and p--p
data of  the BRAHMS--Collaboration\cite{BRAHMSdaupp} at
pseudorapidities of 0, 1, 2.2 and 3.2. In the \textsc{Dpmjet}--III
colculations we are not able to simulate the quite complicated
pseudorapidity acceptances as given in  \cite{BRAHMSdaupp}, instead we
use in all cases a pseudorapidity band of width 0.2 centered at the
nominal pseudorapidity. This difference might be responsable for part of
the differences we find between the model and the data.
There is no essential difference between the agreements for d--Au  and
p--p. The model follows the change of the shapes of the data with
changing pseudorapidity, but the agreement is not perfect. 

\section{ 
p--p collisions in 
{\sc Dpmjet}-III
 }
       
We are here mainly concerned with d--Au collisions, but the changes of
Dpmjet described in the last Section will also change hadron 
production in
p--p collisions. Therefore, let us shortly discuss p--p collisions at
the energy of the RHIC collider.

In Fig.\ref{fig:etacmdau200} we did already compare
\textsc{Dpmjet}-III  with the preliminary data of the 
PHOBOS Collaboration\cite{PHOBOSdaupp} for the pseudorapidity 
distribution of charged hadrons
in p--p collisions  at  $\sqrt{s}$ = 200 GeV.
We found an excellent
agreement similar to the agreement found in earlier comparisons with
data from the CERN--SPS and the TEVATRON colliders.

In Figs. \ref{fig:daucha0pt} 
%\ref{fig:daucha1pt},\ref{fig:daucha2p2pt}, 
and \ref{fig:daucha3p2pt} 
we compared the modified
\textsc{Dpmjet}--III as described in this paper also with the  p--p
preliminary data of  the BRAHMS--Collaboration\cite{BRAHMSdaupp} on charged hadron
transverse momentum distributions at
pseudorapidities of 0, 1, 2.2 and 3.2.

 \begin{figure}[thb]
\vspace{-11mm}
\begin{center}
\includegraphics[height=7cm,width=8cm]{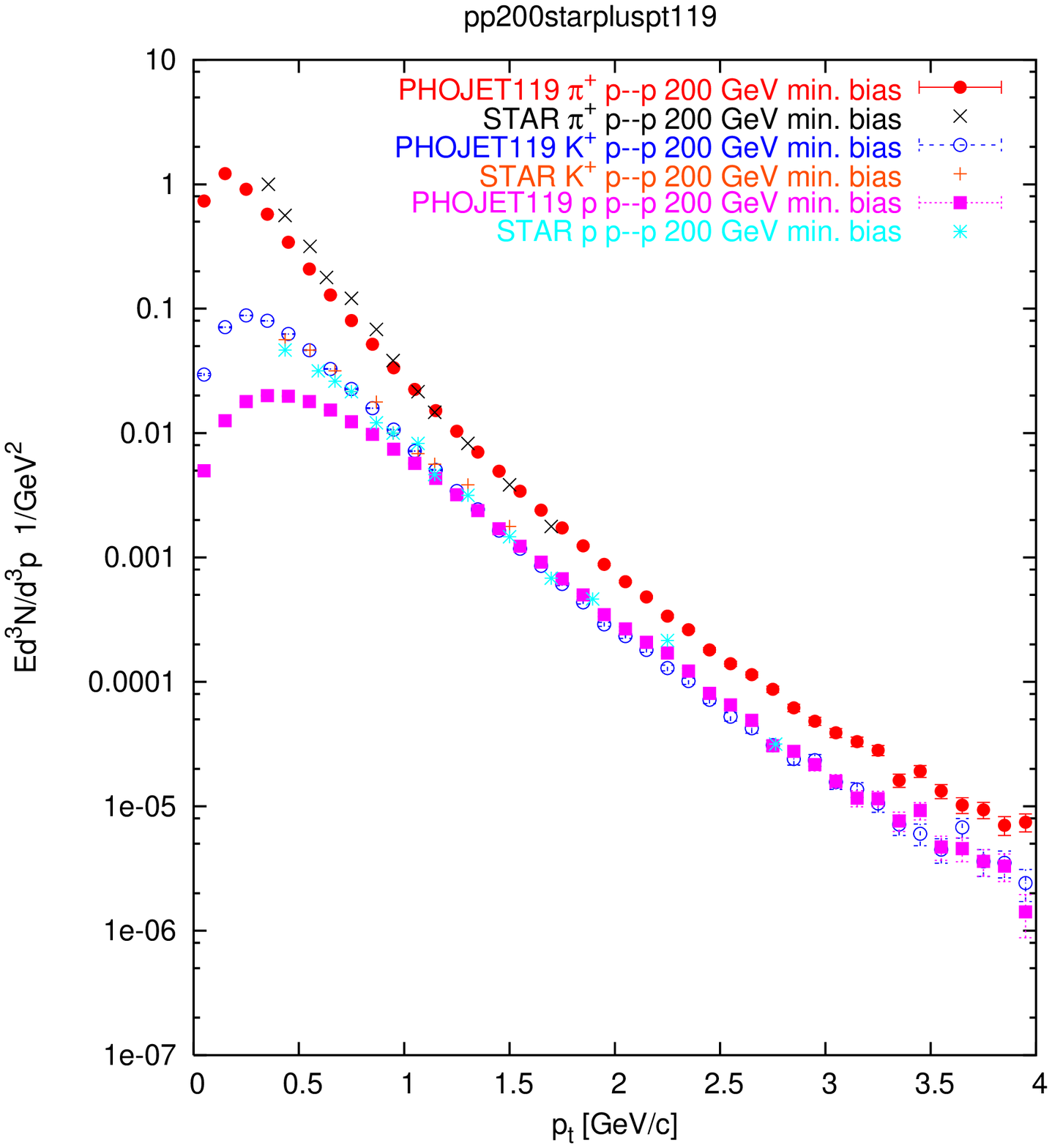}
\end{center}
\vspace{-13mm}

\caption{Transverse momenum distributions of positively charged hadrons
in minimum bias p--p collisions. Compared are the preliminary data from the
STAR--Collaboration\cite{Adams03} to the results of the 
\textsc{Dpmjet}--III with modified transverse momentum distributions in
hadronic soft chain decay as decribed in this paper.
 }

\vspace{-7mm}
\label{fig:ppplusptnew}
\end{figure}

 \begin{figure}[thb]
\vspace{-11mm}
\begin{center}
\includegraphics[height=7cm,width=8cm]{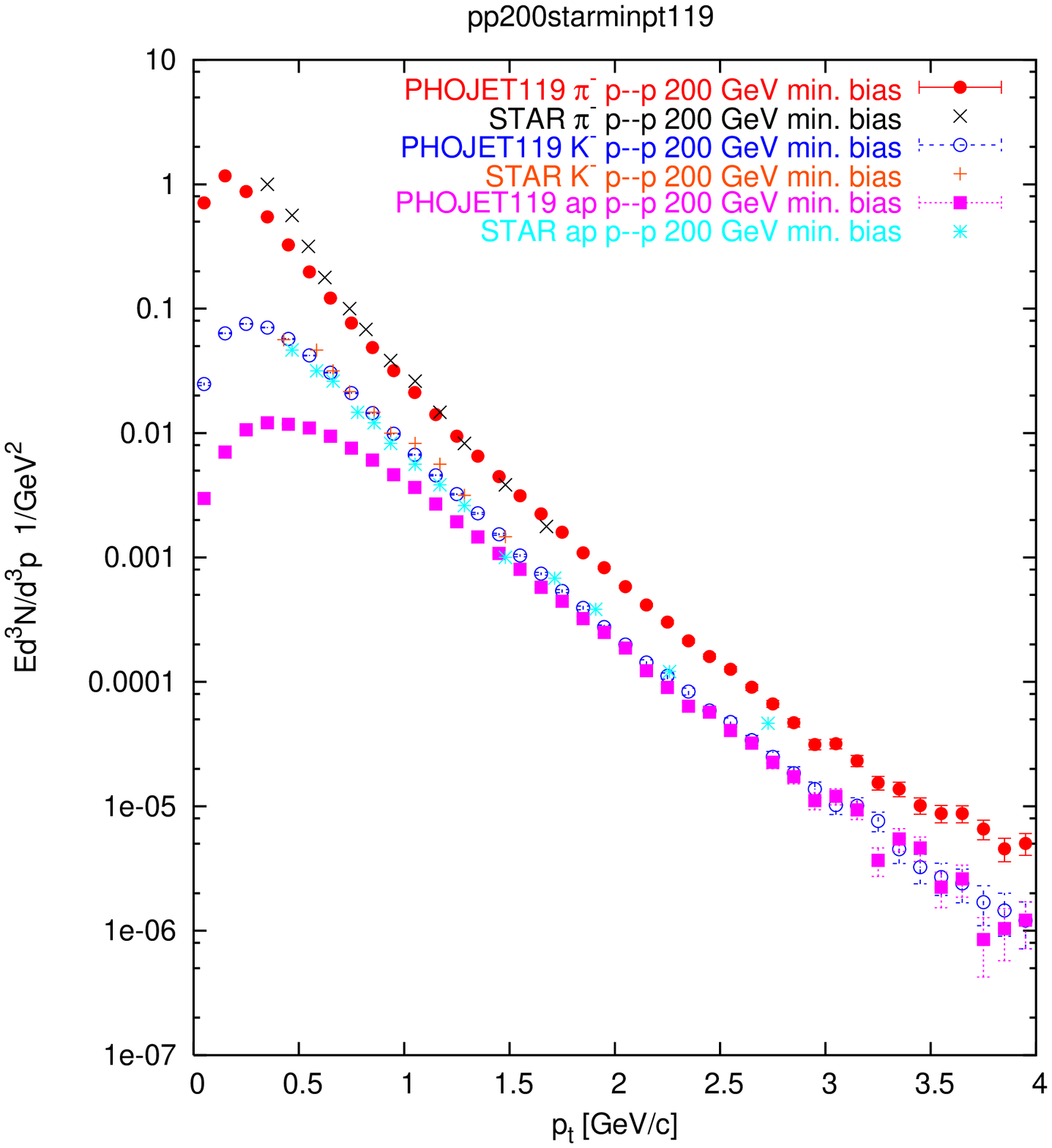}
\end{center}
\vspace{-13mm}

\caption{Transverse momenum distributions of negatively charged hadrons
in minimum bias p--p collisions. Compared are the preliminary data from the
STAR--Collaboration\cite{Adams03} to the results of the 
\textsc{Dpmjet}--III with modified transverse momentum distributions in
hadronic soft chain decay as decribed in this paper.
 }

\vspace{-7mm}
\label{fig:ppminptnew}
\end{figure}

 \begin{figure}[thb]
\vspace{-11mm}
\begin{center}
\includegraphics[height=7cm,width=8cm]{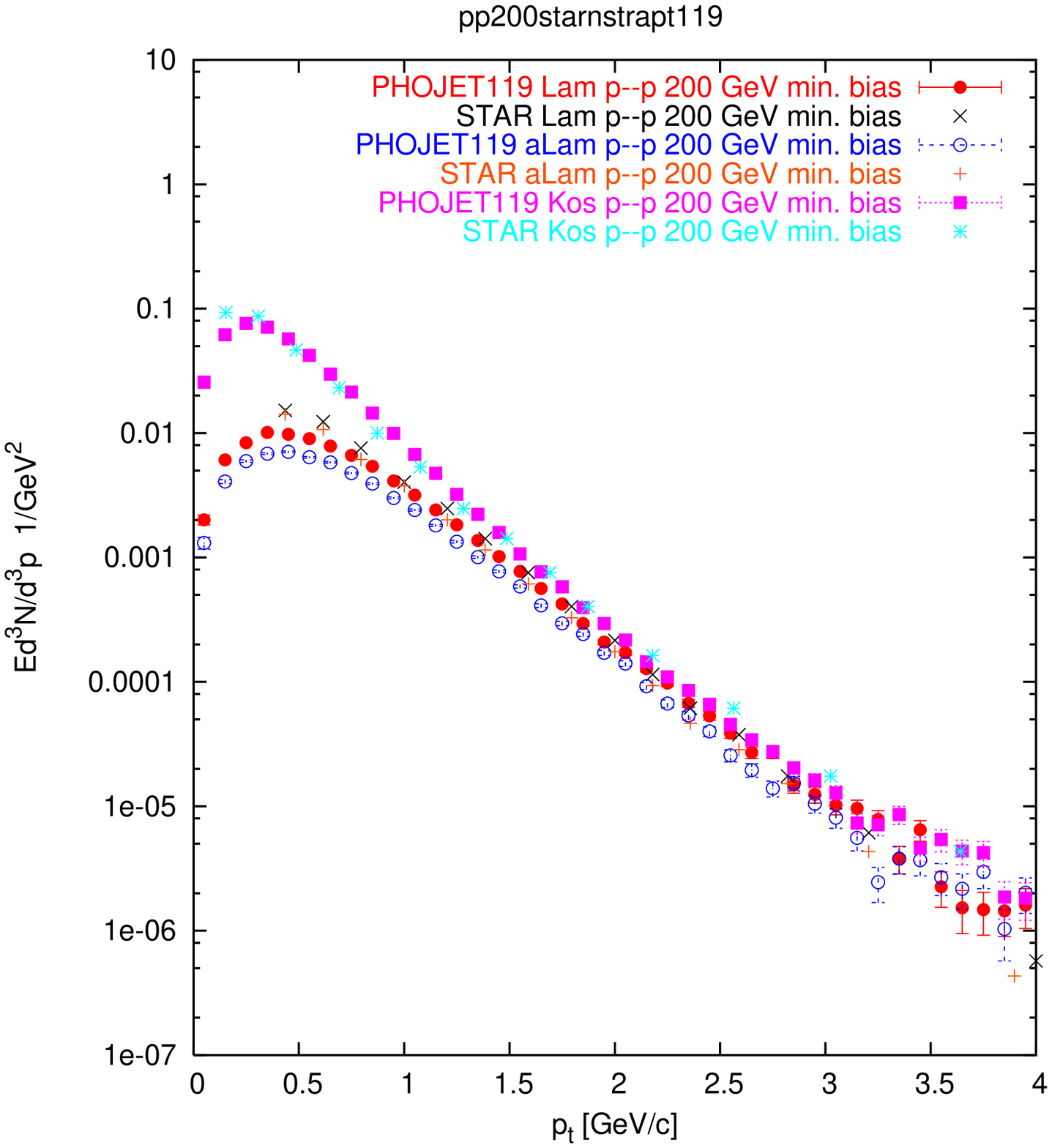}
\end{center}
\vspace{-13mm}

\caption{Transverse momenum distributions of neutral strange hadrons
in minimum bias p--p collisions. Compared are the preliminary data from the
STAR--Collaboration\cite{Adams04} to the results of the 
\textsc{Dpmjet}--III with modified transverse momentum distributions in
hadronic soft chain decay as decribed in this paper.
 }

\vspace{-7mm}
\label{fig:ppnstraptnew}
\end{figure}

Using \textsc{Dpmjet}--III with modified transverse momentum
distributions in
     hadronic soft chain decay in Figs.\ref{fig:ppplusptnew} 
 and  \ref{fig:ppminptnew}  
     we find like in
     d--Au collisions an improved agreement between
     \textsc{Dpmjet}--III and the preliminary data on $\pi^+$, $\pi^-$ K$^+$,
     K$^-$, proton and antiproton transverse momentum distributions
     in p--p collisions from the
          STAR--Collaboration\cite{Adams03}.

Transverse momenum distributions of neutral strange hadrons
in minimum bias p--p collisions were measured by the
STAR--Collaboration\cite{Adams04}.
In Fig.\ref{fig:ppnstraptnew} we compare
\textsc{Dpmjet}--III with modified transverse momentum distributions in
hadronic soft chain decay to these preliminary STAR data. The ageement is
satisfactory.

\section{ Summary 
 }
 
 The data obtained at RHIC are extremely useful to
 improve hadron production models like {\sc Dpmjet}-III.

 Of particular importance are data on hadron production in p--p
 collisions, d-Au collisions and peripheral Au--Au collisions, in all of
 these collisions (unlike central Au-Au collisions) we do not expect any
 change in the reaction mechanism, which might not be accommodated into
 the mechanisms as implemented in {\sc Dpmjet}--III.

 Indeed, comparing {\sc Dpmjet}--III to RHIC 
 data we find three important
 corrections to be applied to {\sc Dpmjet}--III, which otherwise do not
 completely change the independent chain fragmentation model: 
 (i) Percolation and
 fusion of chains, the data from RHIC allow to determine the amount of
 percolation to be implemented into {\sc Dpmjet}--III. (ii) Collision
 scaling of large $p_{\perp}$ hadron production in d--Au collisions: The
 data indicate that we have 
 to change the iteration procedure (of the selection of all soft and
 hard chains in nuclear collisions) in such a way, 
 that collision scaling
 is obtained. (iii) Replacing in soft hadronic collisions the Gaussian
 transverse momentum distribution contained in the {\sc Jetset-- Pythia} code
 \cite{Sjostrand01a} by an exponential distribution.

%%%%%%%%%%%%%%%%%%% bibliography %%%%%%%%%%%%%%%%%%%%%%%%%%%%%%%%%%%
 
 \bibliographystyle{prsty}

 \bibliography{dpm11}
 
%%%%%%%%%%%%%%%%%%%%%%%%%%%%%%%%%%%%%%%%%%%%%%%%%%%%%%%%%%%%%%%

\end{document}